\documentclass[12pt,preprint]{aastex}

\newcommand{\be}{\begin{eqnarray}}
\newcommand{\ee}{\end{eqnarray}}
\newcommand{\eq}[1]{Eq.~(\ref{#1})}
\newcommand{\fig}[1]{Fig.~\ref{#1}}
\newcommand{\pa}{\partial}

\newcommand{\e}[1]{$\times 10^{#1}$}

\title{A Cosmic Ray Positron Anisotropy due to Two Middle-Aged, Nearby Pulsars?}
\shorttitle{Constraints on the lepton content of PWN}
\author{I. B\"usching}
\affil{Unit for Space Physics, North-West University, Potchefstroom Campus, Private Bag X6001, Potchefstroom 2520, South Africa}
\affil{The Centre for High Performance Computing, CSIR Campus, Rosebank, Cape Town, South Africa}
\email{fskib@puk.ac.za}
\author{O. C. de Jager}
\affil{Unit for Space Physics, North-West University, Potchefstroom Campus, Private Bag X6001, Potchefstroom 2520, South Africa}
\affil{The Centre for High Performance Computing, CSIR Campus, Rosebank, Cape Town, South Africa}
\affil{Dept. of Science \& Technology and National Research Foundation Reasearch Chair: Astrophysics \& Space Science}
\email{Okkie.DeJager@nwu.ac.za}
\author{M.S. Potgieter}
\affil{Unit for Space Physics, North-West University, Potchefstroom Campus, Private Bag X6001, Potchefstroom 2520, South Africa}
\affil{The Centre for High Performance Computing, CSIR Campus, Rosebank, Cape Town, South Africa}
\email{marius.potgieter@nwu.ac.za}
\author{C. Venter}
\affil{Unit for Space Physics, North-West University, Potchefstroom Campus, Private Bag X6001, Potchefstroom 2520, South Africa}
\affil{The Centre for High Performance Computing, CSIR Campus, Rosebank, Cape Town, South Africa}
\email{fskcv@puk.ac.za}

\begin{abstract}
Geminga and B0656+14 are the closest pulsars with characteristic ages in the range of 100\,kyr to 1\,Myr. 
They both have spindown powers of the order 3\e{34} erg/s at present.
The winds of these  pulsars had most probably powered  pulsar wind nebulae (PWNe) that  broke up less than about 100\,kyr after the birth of the pulsars. 
Assuming that leptonic particles accelerated by the pulsars  were confined in the PWNe and were released into the interstellar medium (ISM) on breakup of the PWNe, we show that, depending on the pulsar parameters, both pulsars make a non-negligible contribution to the local cosmic ray (CR) positron spectrum, and they may be the main contributors above several GeV. The relatively small angular distance between  Geminga and B0656+14 thus implies an anisotropy in the local CR positron flux at these energies.
We calculate the contribution of these pulsars to the locally observed CR electron and positron spectra depending on the pulsar birth period and the magnitude of the local CR diffusion coefficient. We further give an estimate of the expected anisotropy in the local CR positron flux.
Our calculations show that within the framework of our model, the local CR positron spectrum imposes constraints on pulsar parameters for Geminga and B0656+14, notably the pulsar period at birth, and also the local interstellar diffusion coefficient for CR leptons.
\end{abstract}

\keywords{pulsars: individual (Geminga,  B0656+14), acceleration of particles, cosmic rays, diffusion }

\begin{document}
\maketitle

\section{Introduction}
Geminga and B0656+14 are the closest pulsars with characteristic ages in the range of 100\,kyr to 1\,Myr \citep{atnf}. They both have spindown powers of the order 3\e{34} erg/s at present.
The winds of these  pulsars had most probably powered  pulsar wind nebulae (PWNe) that broke up less than about 100\,kyr after the birth of the pulsars. The reason for this statement is that we do not observe PWNe associated with pulsars older than 100\,kyr.

Assuming that leptonic particles accelerated by the pulsars prior to breakup were confined in the PWNe and were released into the interstellar medium (ISM) on breakup of the PWNe (i.e. when the pressure of the ambient ISM exceeds the magnetic pressure $B^2/(8\pi)$ in the PWN), we calculate the contribution of these particles to the locally observed cosmic ray (CR) electron and positron spectra. 
We further calculate the expected anisotropy in the positron local interstellar spectrum (LIS) in the case of energy-dependent diffusion.  

\section{Positrons from PWNe}
We discuss the acceleration of particles by pulsars in the framework of the polar cap (PC) model \citep[see e.g.\ the review of][]{2004AdSpR..33..552B}.
Given the relatively high surface magnetic fields of $B_s=1.6$e12\,G and $B_s=4.7$e12\,G for Geminga and B0656+14 respectively, a single primary electron released from the stellar surface will induce a cascade of electron-positron pairs in the magnetospheres of these pulsars. This amplification is modeled by introducing a multiplicity $M^\prime$.
The flux of primary electrons from the pulsar PC is given by the Goldreich-Julian current:
\begin{equation}
I_{\rm GJ} \approx 2c\rho_{\rm GJ}A_{\rm PC} \approx \frac{B_{\rm s}\Omega^2R^3}{c} = \sqrt{6c{L}_{\rm sd}},
\end{equation}
with $\rho_{\rm GJ}$ the Goldreich-Julian charge density \citep{GJ69}, and $A_{\rm PC} \approx \pi R_{\rm PC}^2$ the PC area. 

We assume that the electrons and positrons from the pulsar are re-accelerated at the pulsar shock, and model the particle spectrum by a power law with spectral index of 2, and with a maximum energy of \citep{venterdejager06}
\be
E_{\rm max}&=&\epsilon e \kappa \sqrt{\left(\frac{\sigma}{\sigma+1}\right)\frac{ L_{\rm sd}}{c}},
\ee
where $\kappa$ is the compression ratio at the shock, $L_{\rm sd}$ the spindown power, and $\sigma$ the magnetization parameter \citep{1984ApJ...283..694K}. This maximum energy stems from a condition on particle confinement: we require that the ratio between the Lamor radius $r_L$ and the radius  $r_S$ of the pulsar shock should be less than unity. 
We assume a maximum ratio $\epsilon=r_L/r_S=0.1$ and that for larger Lamor radii, the curvature of the shock results in particle losses. For the calculations presented in this paper, we assume $\kappa\,=\,3$.
As observations suggest that $\sigma$ depends on the age of the pulsar ($\sigma=0.003$ is found for the Crab pulsar \citep{1984ApJ...283..694K}, which has an age of 1\,kyr, but  $\sigma=0.1$ for 11\,kyr Vela \citep{2003ApJ...593.1013S}), we thus assume 
\be
\sigma(t)&=&\sigma_0\left(\frac{t}{1\,{\rm kyr}}\right)^{3/2}, 
\ee
with $\sigma_0=0.003$.
The fraction of the spindown power deposited in particles can be written in terms of the magnetization parameter  $\sigma$ \citep{2006cosp...36.3071B}:
\be
\eta_{\rm part}&=&f\,\frac{1}{1+\sigma}.
\label{etapart}
\ee
Here, we introduce a geometry factor $f=\Omega_{acc}/(4\pi)$, as 
$\sigma$ is supposed to be significantly less than unity \citep{1984ApJ...283..694K}, implying a large $\eta_{\rm part}$. Without $f$, from Eq.~\ref{etapart} $\eta_{\rm part}$ is of the order unity for $\sigma \ll 1$. This however is in contradiction with observations, indicating $\eta_{\rm part}\ll 1$, as found e.g. for the Vela PWN \citep{2003ApJ...593.1013S}. 
We adopt $f=1/2$. 

At any time the conditions
\be
\int_{E_{\rm min}}^{E_{\rm max}} Q^\prime(E,t)dE & = & \frac{M^\prime I_{\rm GJ}}{e}\label{eq:current}\\
\int_{E_{\rm min}}^{E_{\rm max}} Q^\prime(E,t)EdE & = & \eta_{\rm part}L_{\rm sd}(t)\label{eq:energy}
\ee
have to hold \citep{2003ApJ...593.1013S}, where $ Q^\prime(E,t)=K^\prime E^{-2}$ is the assumed particle spectrum at the pulsar wind shock. 
${E_{\rm min}}$ is assumed to be similar to  the inferred value of the Crab. A value of   100~MeV is adopted.

For a non-decaying pulsar magnetic field (i.e. $\dot{P}P^{n-2} = \dot{P}_0P_0^{n-2}$), the time-evolution of $L_{\rm sd}$ is given by \cite{1974MNRAS.167....1R}
\be
L_{\rm sd}(t) &=& L_{\rm sd,0}\left(1+\frac{t}{\tau_0}\right)^{-\frac{n+1}{n-1}}\!\!\!\!\!\!\!\!\!\!\!\!, \label{eq:Lsd}
\ee
with $n = 3$ representing a dipolar magnetic field, and $\tau_0 = P_0/((n-1)\dot{P}_0)$, $P_0$ the birth period, and $\dot{P}_0$ the period's time-derivative at pulsar birth. (The subscript `0' denotes quantities at pulsar birth). These quantities are connected to the spindown luminosity at birth via
\begin{equation}
L_{\rm sd,0} =\left| -\frac{4\pi^2I\dot{P}_0}{P_0^3}\right|.
\label{eq:Lsd0}
\end{equation}
We assume that the particles are confined in the PWN for a time $T$, after which the PWN breaks up and releases them into the surrounding ISM. The time evolution of the particle spectrum $Q(E,t)$ inside the PWN is described by
\be
\frac{\partial Q(E,t)}{\partial t}-  Q^\prime(E,t)=\frac{\partial}{\partial E}\left(B_{\rm PWN}(t)^2E^2Q(E,t)\right),
\label{pwnevo}
\ee
where we assumed a  decaying  magnetic field in the PWN
\be
B_{\rm PWN}(t)&=&\frac{1200}{\left(1+t/{\rm kyr}\right)^2}\,\,\, [{\rm \mu G}]. 
\ee
This parametrization for $B$ is justified as follows. 
After 10 to 20\,kyr, the PWN field strength is already of the order of $5\,\mu$G 
as observed by HESS from a number of PWN \citep[see e.g.][]{2008arXiv0803.2104D}
and after $\sim 100$\,kyr we expect that the ISM pressure will randomize the PWN field structure, 
leading to relatively fast diffusive escape of charged particles from the nebula.   
Although the actual age for breakup is difficult to estimate, we assume 
to a first order a number of less than 100\,kyr. GLAST observations of a limiting age for mature PWN 
may shed more light on this epoch of breakup \citep{2008arXiv0803.2104D} .

Eq. \ref{pwnevo} can be solved using the Green's function formalism. The particle spectrum at time $T$  is given by
\be
Q(E,T)&=&\int_0^{T}Q^{\prime}(E_0,t_0)E_0^2E^{-2}\Theta\left(E_0-E_{\rm min}\right)\Theta\left(E_{\rm max}-E_0\right)dt_0,
\label{geminga:spec}
\ee
where $\Theta$ is the Heaviside step function and 
\be
E_0&=&\frac{E}{E\int_t^{t_0}B_{\rm PWN}(t^{\prime})^2dt^{\prime}+1}.
\ee
Our model predicts $\approx$equal numbers of positrons and electrons to be accelerated by pulsars, thus Eq.~\ref{geminga:spec} also describes the source function for CR electrons.  
\section{Propagation of CR positrons and local anisotropy}
\begin{figure*}[th]
  \hfill
  \begin{minipage}[t]{.45\textwidth}
    \begin{center}  
      \includegraphics[width=\linewidth]{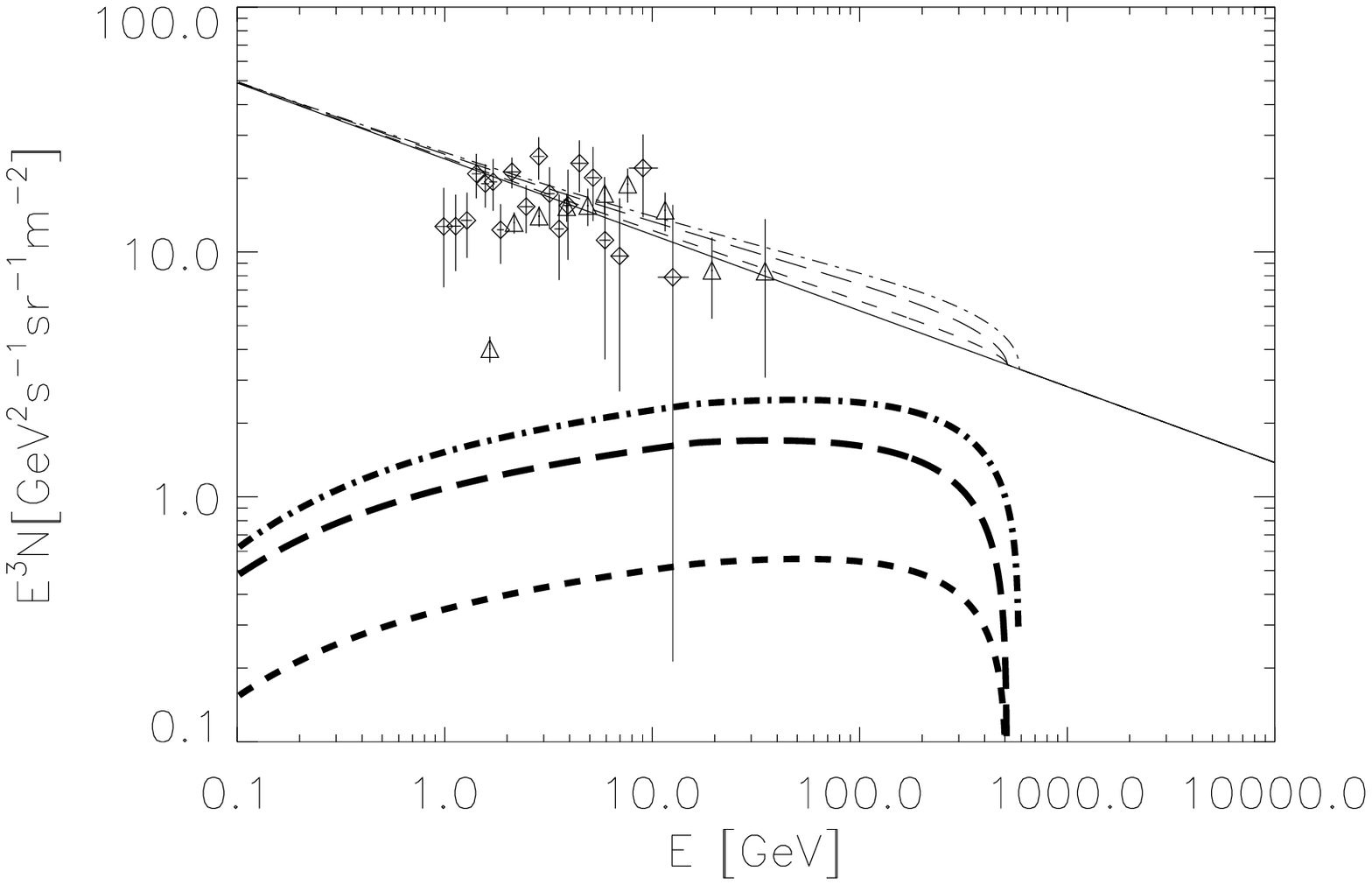}
    \end{center}
  \end{minipage}
  \hfill
  \begin{minipage}[t]{.45\textwidth}
    \begin{center}  
      \includegraphics[width=\linewidth]{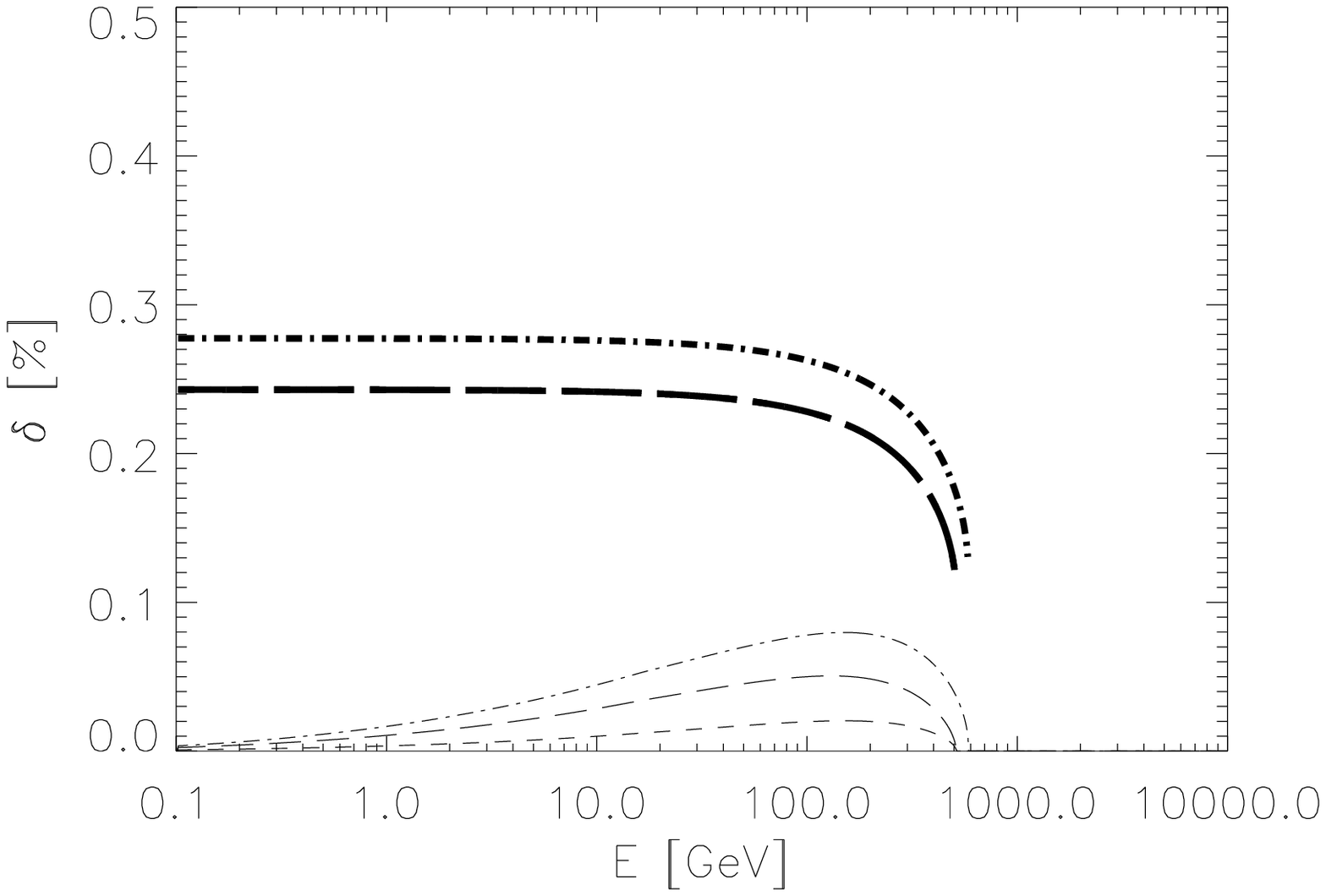}
    \end{center}
  \end{minipage}
  \hfill
\caption{Left panel: Contribution of Geminga to the positron LIS for $k_0=0.1\,$kpc$^2$\,Myr$^{-1}$ and $P_0\,=\,40\,$ms, $T=20\,$kyr (long dashed line), $P_0\,=\,40\,$ms, $T=60\,$kyr (dot-dashed line), and $P_0\,=\,60\,$ms, $T=20\,$kyr (dashed) on top of an isotropic background (solid line). The thin lines mark the combined spectra (pulsar contribution plus background), whereas the thick lines give the contribution of the pulsar alone. Also shown are data from \citet{2000ApJ...532..653B} (diamonds) and \citet{2001ApJ...559..296D} (triangles).
Right panel: the expected local anisotropy in case only Geminga contributes to the LIS (thick lines), and in case Geminga contributes on top of an isotropic background positron flux (thin lines) as given by \citet{1998ApJ...498..779B} (solid line in left panel). The line styles correspond to the cases as given for the left panel. 
The thick dashed and long dashed lines coincide.}
\label{fig:geminga}
\end{figure*}

\begin{figure*}[th]
  \hfill
  \begin{minipage}[t]{.45\textwidth}
    \begin{center}  
      \includegraphics[width=\linewidth]{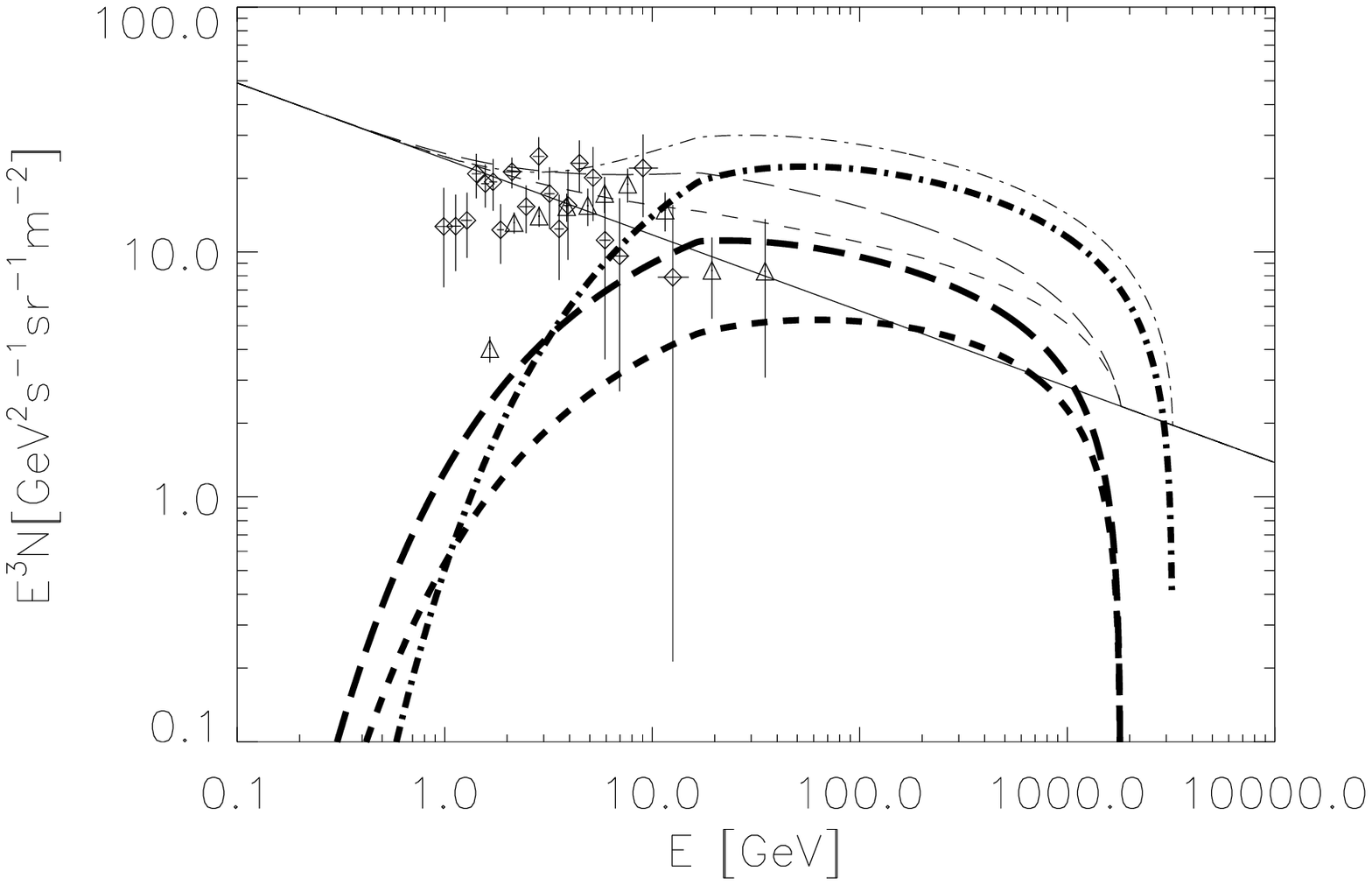}
    \end{center}
  \end{minipage}
  \hfill
  \begin{minipage}[t]{.45\textwidth}
    \begin{center}  
      \includegraphics[width=\linewidth]{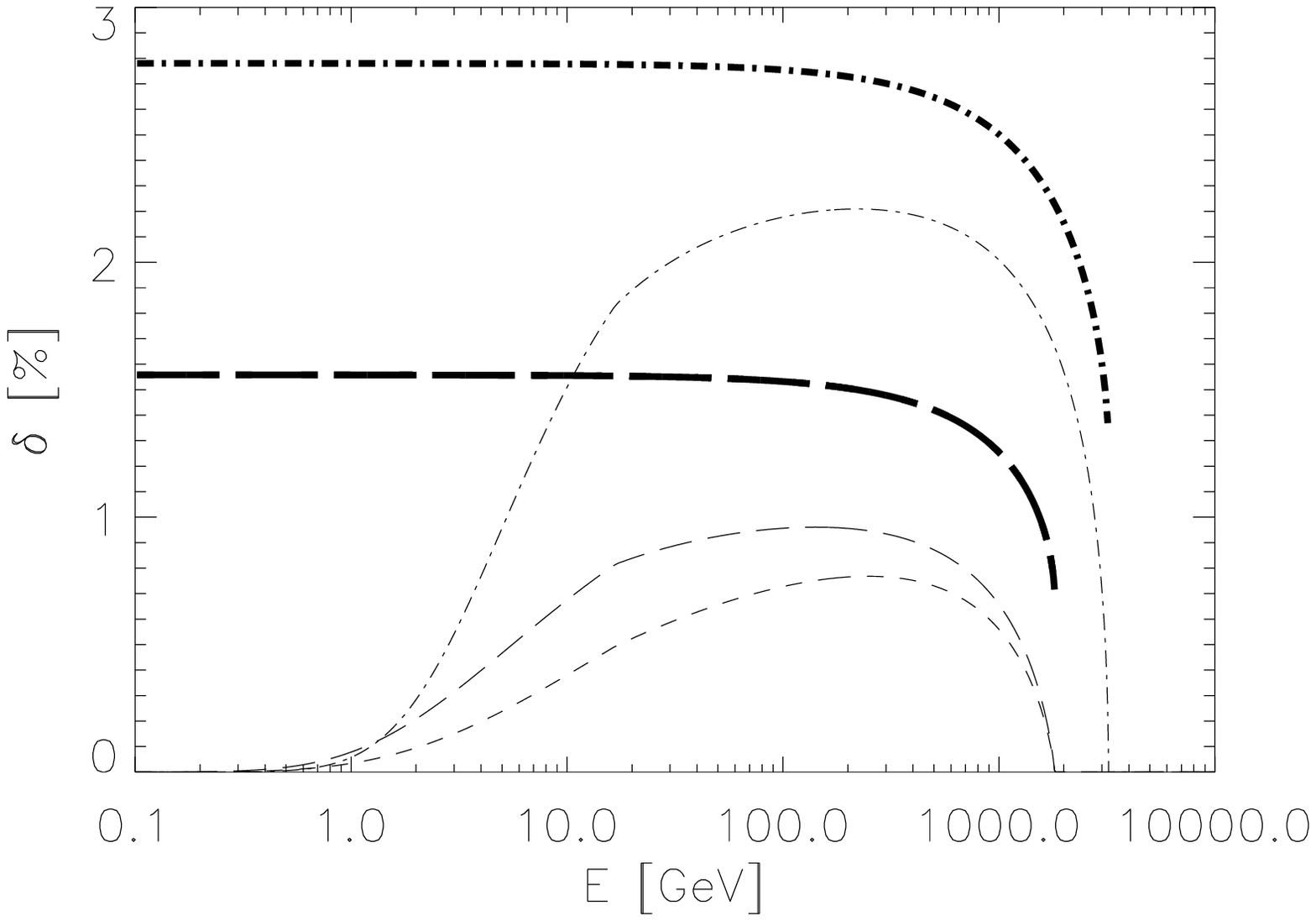}
    \end{center}
  \end{minipage}
  \hfill
\caption{Same as \fig{fig:geminga} but for B0656+14.}
\label{fig:bo656}
\end{figure*}

The propagation of CR electrons and positrons in case of the diffusion coefficient $k$ being spatially constant, is described by 
\be
\frac{\pa N}{\pa t} -S &=& k\Delta N -\frac{\pa }{\pa E}\left(b\,N\right),
\label{prop:timedep}
\ee
where $N$ is the differential number density, $S$ the source term, $\Delta$ the Lapace operator, and $b$ the 
rate of energy losses. 
For a functional form of the diffusion coefficient $k=k_0E^{\alpha}$, with $\alpha\,=\,3/5$ and  the energy losses
$b=b_0E^2$  (i.e.\ synchrotron and inverse Compton losses) one can find a 
 Green's function solving \eq{prop:timedep} in the literature \citep{1990acr..book.....B}.
It is given by
\be
G(\vec{r},\vec{r_0},E,E_0,t,t_0)&=& \delta\left(t-t_0-\frac{1}{b_0}\left( E_0^{-1}-E^{-1}\right) \right)\frac{\exp\left(-\frac{\left(\vec{r}-\vec{r_0}\right)^2}{\lambda}\right)}{b(\pi\lambda)^{1.5}}, 
\ee
with
\be
\lambda&=& 4\frac{k_0\left(E^{\alpha -1}-E_0^{\alpha-1}\right)}{\left(1-\alpha\right)b_0}.
\ee
%
%
Thus, the solution of \eq{prop:timedep}
is given by
\be
N&=&\int \int \int G(\vec{r},\vec{r_0},E,E_0,t,t_0)Q(E_0)\delta\left(t_0-t_i\right)\delta\left(\vec{r_0}-\vec{r_i} \right)dE_0dt_0d\vec{r_0},
 \label{eq:spec}
\ee
where $Q(E_0)$ is given by \eq{geminga:spec}, $\vec{r_i}$ and $t_i$ are the place and time of injection, repectively.
We note, that the energy $E$ of a particle at time $t$ is linked to its energy $E_0$ at injection by 
\be
E_0={E}/\left({\left(t-t_i\right)b_0E+1}\right).
\label{eq:e0}
\ee
The anisotropy in the CR flux can be calculated in the context of diffusion as \citep{1964ocr..book.....G}
\be
\delta=\frac{I_{\rm max}-I_{\rm min}}{I_{\rm max}+I_{\rm min}}=\frac{3\,k\left|\nabla N\right|}{c\,N},
\label{eq:delta}
\ee
where $\nabla N$ denotes the gradient of $N$.
The expected anisotropy in the positron LIS was calculated assuming the contribution of a nearby source, as given by \eq{eq:spec}, on top of an isotropic background. For the background we assumed a power law fit given by \citet{1998ApJ...498..779B}. The calculated anisotropies are given in the right panels of Figs.~\ref{fig:geminga} and  \ref{fig:bo656} (thin lines).

To get an estimate of the maximum expected anisotropy, we also calculate the anisotropy assuming that the whole CR positron flux originates from a point source (thick lines in the right panels of Figs.~\ref{fig:geminga} and  \ref{fig:bo656}). For energy-independent diffusion, \citet{1972ChJPh..10...16M} derived the simple relation
\be
\delta=\frac{3}{2c}\frac{r_i}{t_i}.
\label{eq:shen}
\ee 
Allowing for energy-dependent diffusion, we get, inserting \eq{eq:spec} into \eq{eq:delta},
\be
\delta=\frac{3}{2\,c}{ r_i}\,{ b_0}\, \left(\alpha-1  \right) {E}^{{\alpha}} \left( {E}^{\alpha -1}- E_0 ^{\alpha-1} \right) ^{-1}.
\label{eq:ingo}
\ee 
As in the case of \eq{eq:shen}, \eq{eq:ingo} also does not depend on the magnitude of the diffusion coefficient, only on the distance $r_i$ to the source, and via \eq{eq:e0}, on the time since the injection of the particles into the ISM. In the limit of $E\rightarrow 0$, \eq{eq:ingo} reduces to \eq{eq:shen}. 

We calculated the contribution from Geminga and B0656+14 to the positron LIS for distances of 157\,pc \citep{C96} and 290\,pc \citep{atnf} respectively (assuming $T\,=\,20\,$kyr and $T\,=\,60\,$kyr), in addition to the expected anisotropies in the positron LIS. The results for birth periods of 40\,ms and 60\,ms are plotted in the left panels of Figs.~\ref{fig:geminga} and \ref{fig:bo656}, where we compare our calculations with the measurements from \citet{2000ApJ...532..653B} and \citet{2001ApJ...559..296D}.

\section{Conclusions}
We have shown that one can expect a non-negligible CR positron component in the LIS from nearby pulsars that may become dominant above several GeV, in agreement with \citet{1995PhRvD..52.3265A} who showed that the high-energy positron LIS  may be explained by a young, nearby source. In the context of our model, we are able to constrain the permissible pulsar birth period $P_0$, depending on the magnitude of the interstellar diffusion coefficient. For the two nearest pulsars with characteristic ages in the range 1e5\,yr to 1e6\,yr, Geminga and B0656+14, we show that in particular for B0656+14 one can expect, in the absence of a background flux, an anisotropy in the positron LIS of up to almost 3\%, significantly larger than the expected value of $\approx$0.25\% for Geminga. As shown in Figs.~\ref{fig:geminga} and \ref{fig:bo656}, the observed anisotropy also gives an estimate of the contribution of the pulsar to the positron LIS. On the other hand, a measured anisotropy larger than the $\approx$3\% we obtained for B0656+14 would be, from \eq{eq:shen} and \eq{eq:ingo}, an indication of the existence of an even younger, nearby source, e.g.\ a longer lifetime $T$ of the PWN.
We also note that the predicted flux from PSR B0656+14 appears to overpredict the observed flux above 10\,GeV. This implies more severe constraints on the pulsar output, whereas Geminga's parameters are not that severely constrained by CR positron observations.

We remark that Galactic CR, including electrons and positrons, are subjected to solar modulation at energies below $\approx$10\,GeV. The encounter of these particles with the solar wind and imbedded magnetic field causes a heliospheric anisotropy that is primarily determined by the combined modulation effects of convection, diffusion, and drifts - all solar cycle dependent. Drifts will cause this anisotropy to have a 22-year cycle. CR electrons and positrons at 1\,GeV at Earth may therefore exhibit a heliospheric anisotropy of up to a few percent, assuming that they enter the heliosphere isotropically \citep{2004ApJ...602..993P}.
%
%
It will be an interesting exercise to determine how this anisotropy will change if the LIS is anisotropic. 
However, the anisotropy that we predict here is the largest above 10\,GeV, an energy range at which only 
the PAMELA mission \citep{2004NuPhS.134...39B} may be able to gather
sufficient statistics to find an anisotropy of the predicted magnitude.

\section{Acknowledgments}
This work is supported by the SA National Research Foundation
and the SA Centre for High Performance Computing.

%
\end{document}